\documentclass[12pt]{iopart}
\usepackage{xcolor,epsfig,bm}

\newcommand{\ov}{\overline}

\newif\ifredline

\redlinefalse

\begin{document}

\frenchspacing
\title{Insights from number theory into the critical Kauffman model with connectivity one}
\author{F. C. Sheldon and T. M. A. Fink}
%\affiliation{London Institute for Mathematical Sciences, Royal Institution, 21 Albemarle St, London W1S 4BS, UK}

\date{\today}
\begin{abstract}
\noindent
{\ifredline\color{red}\else\color{black}\fi
The Kauffman model of genetic computation highlights the importance of criticality at the border of order and chaos.
The model with connectivity one is of special interest because it is exactly solvable.
But our understanding of its behavior is incomplete, and much of what we do know relies on heuristic arguments.
Here, we show that the key quantities in the model are intimately related to aspects of number theory.
Using these links, we derive improved bounds for the number of attractors as well as the mean attractor length, which is harder to compute.
Our work suggests that number theory is the natural language for deducing many properties of the critical Kauffman model with connectivity one, and opens the door to further insight into this deceptively simple model.
}
\end{abstract}

\maketitle
\noindent
% - - - - - - - - - - - - - - - - - - - - - - - - - - - - - - - - - - - - - - - - - - - - - - - -
\noindent {\sf\textbf{\textcolor{black}{\large 1 Introduction}}}
% - - - - - - - - - - - - - - - - - - - - - - - - - - - - - - - - - - - - - - - - - - - - - - - -
\\ \noindent
% \emph{\textcolor{red}{Overture.}}
The Kauffman model is the simplest and most-studied model of genetic computation. %\cite{Flyvbjerg88,Drossel05,Fink23PRE,Fink23PRL,Socolar03,Troein03,Peixoto10}.
Originally introduced as a toy model of cell type \cite{Kauffman69a,Kauffman69b}, it has since garnered interest from many statistical physicists as an archetypal disordered system \cite{Aldana03,Drossel08}.
In particular, it displays a critical regime separating frozen and chaotic dynamics, at which many physical and biological systems seem poised~\cite{Munoz18,Daniels18}.
{\ifredline\color{red}\else\color{black}\fi
In this regime, the model exhibits a dynamical phase transition~\cite{Derrida86a} and a disordered landscape similar to the mean-field spin glass~\cite{Derrida86b}.
The Kauffman model is thus a bridge between deterministic non-equilibrium systems and equilibrium statistical mechanics.
But while the physics of the mean-field spin glass is well understood, a similar understanding of disordered Boolean networks has remained elusive.
}
% - - - - - - - - - - - - - - - - - - - - - - - - - - - - - - - - - - - - - - - - - - - - - - - -
\\ \indent
% \emph{\textcolor{red}{Definition.}}
{\ifredline\color{red}\else\color{black}\fi
The Kauffman model is a discrete dynamical system on a directed network of $N$ nodes.
The network is generated randomly such that for each node $K$ inputs are drawn uniformly from the pool of all nodes.
As a result, each node has $K$ inputs but can have any number of outputs.
A typical network for $N=35$ and $K=1$ is shown in Fig. 1.
\\ \indent
The state of each node is 0 or 1 and is a Boolean function of the states of its $K$ inputs.
The Boolean functions drawn randomly from the $2^{2^K}$ functions on $K$ inputs.
The distribution over these is usually uniform for the 16 possible functions for $K=2$ or the four (but sometimes just two) for $K=1$.
Once chosen, the network and Boolean functions do not change as the states are updated.
At each time step, the states are updated simultaneously.
}
% - - - - - - - - - - - - - - - - - - - - - - - - - - - - - - - - - - - - - - - - - - - - - - - -
\\ \indent
% \emph{\textcolor{red}{Criticality.}}
Because the number of states of the network is finite, the dynamics will eventually fall into a repeating sequence of states called an attractor or cycle---we use these terms interchangeably.
Depending on the number of inputs $K$ and the distribution of Boolean functions, the long-term behavior of the Kauffman model falls into two regimes~\cite{Aldana03,Drossel08}.
In the frozen regime, attractor lengths die out with the size of the network and small differences in an initial state do not grow.
In the chaotic regime, attractor lengths grow with $N$ and small differences in an initial state are amplified, leading initially close trajectories to diverge.
These two regimes are separated by a critical boundary, where a perturbation to one node propagates to, on average, one other node.
%{\color{black}
%Surprisingly, for $K=2$ the critical boundary occurs when the distribution over Boolean functions is uniform.
%}
% - - - - - - - - - - - - - - - - - - - - - - - - - - - - - - - - - - - - - - - - - - - - - - - -
\\ \indent
%\emph{\textcolor{red}{$K=2$ scaling.}}
For over half a century, scientists have sought to understand the number and length of attractors in the critical Kauffman model. 
In the original $K=2$ model, numerical evidence led Kauffman to propose that the number of attractors grows as $\sqrt{N}$~\cite{Kauffman69b}, which he and Socolar later revised to faster than linear~\cite{Socolar03}.
Bastolla and Parisi suggested a stretched exponential based on numerical and scaling arguments~\cite{Bastolla98a,Bastolla98b}.
Then, in an elegant analytic study, Samuelsson and Troein~\cite{Troein03} proved that the number of attractors grows faster than any power law.
% - - - - - - - - - - - - - - - - - - - - - - - - - - - - - - - - - - - - - - - - - - - - - - - -
\\ \indent
%\emph{\textcolor{red}{$K=1$ scaling.}}
The intense effort to solve the $K=2$ model sparked interest in the model with $K=1$.
Here, the deceptively simple network is composed of loops with trees branching off of them (Fig. 1).
Because the nodes in the trees are slaves to the loops, they do not contribute to the number or length of attractors, which are set solely by the $m$ nodes in the loops.
In the critical version of the model, all of the Boolean functions must be copy or invert~\cite{Flyvbjerg88, Drossel05}.
In this regime, Flyvbjerg and Kj\ae r found that the growth rate of the number of attractors is at least $2^{0.63 \sqrt{N}}$~\cite{Flyvbjerg88}.
Using a simpler calculation, Drossel \emph{et al.} obtained a slightly slower growth rate of $2^{0.59 \sqrt{N}}$~\cite{Drossel05}.
% - - - - - - - - - - - - - - - - - - - - - - - - - - - - - - - - - - - - - - - - - - - - - - - -
\\ \indent
{\ifredline\color{red}\else\color{black}\fi 
The combination of the loop structure for $K=1$ and simultaneous updates creates an unexpected connection with number theory.
Here the number and length of attractors depends on the divisibility of the loops.
Leveraging this allows us to translate bounds from number theory to bounds on the Kauffman model's dynamics.
%Number theory seems to lie at the heart of the $K=1$ Kauffman model.
%In our concluding remarks, we discuss the scope and limitations of this connection and point to promising directions for future research.
%As the tools we borrow from number theory are unfamiliar in the physics community, we devote a series of subsections to introducing their core ideas.
}
\\ \indent
%\emph{\textcolor{purple}{This paper.}}
This paper is organized as follows.
In part 2 we describe the $K=1$ Kauffman model and 
introduce the two key quantities that characterize it: the number of attractors and the mean attractor length.
% Part 3
In part 3 we show that the number of attractors is at least $2^{1.25 \sqrt{N}}$.
Our approach makes use of Landau's function---the maximum value of the least common multiple of the partitions of a number---and we give an exposition of it here.
% Part 4
In part 4, we show that the mean attractor length is at least $2^{\sqrt{\epsilon N}/\ln \sqrt{\epsilon N}}$, where $\epsilon$ is a small constant.
Our approach makes use of a generalization of Merton's third theorem, which evaluates the product of $1-1/p$ over all primes $p$.
% Part 5
In the discussion in part 5 we compare our results for the number and length of attractors with other known analytic results.
We also discuss why number theory is the right language for understanding our model and consider some open questions for further research. 
% - - - - - - - - - - - - - - - - - - - - - - - - - - - - - - - - - - - - - - - - - - - - - - - -
\begin{figure}[b!]
\noindent
\begin{center}
      \epsfxsize=0.92 \columnwidth \epsfbox{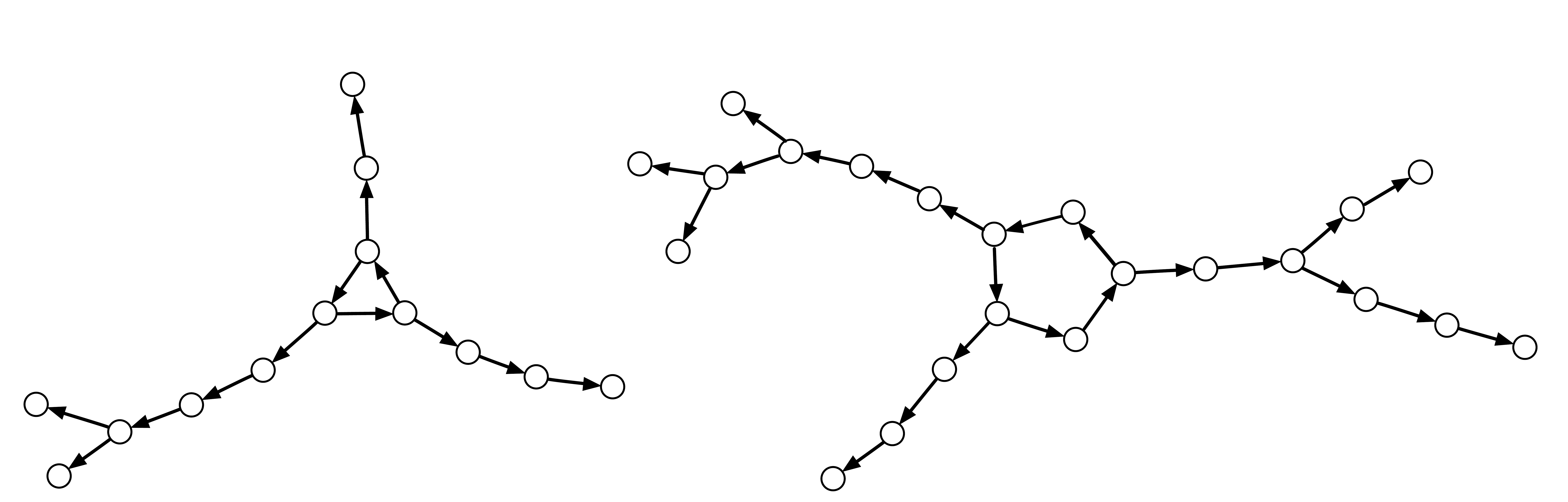} \\
      \epsfxsize = 0.70 \columnwidth \epsfbox{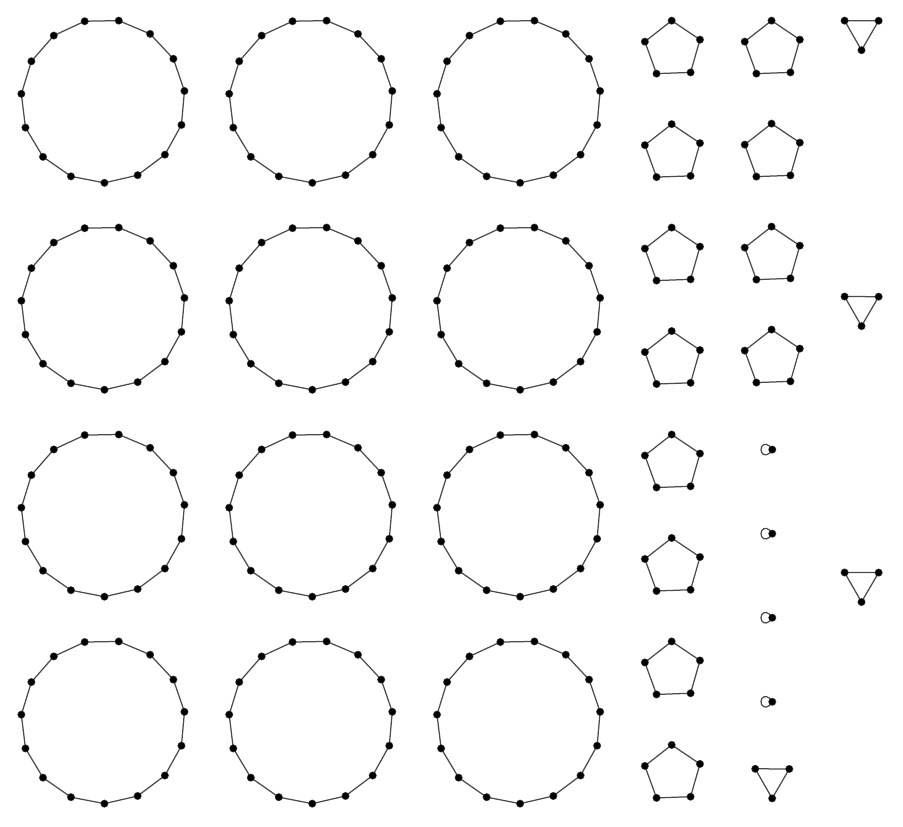}
\end{center}
\begin{small}
\caption{
{\bf A Kauffman network and its attractors.} 
{\bf Top.} 
This typical Kauffman network with connectivity one has $N=35$ nodes and $m=8$ nodes in loops, namely, a 3-loop and a 5-loop.
Only the nodes in loops contribute to the number and length of attractors.
{\bf Bottom.} 
When all of the Boolean functions are copy, the $2^8 = 256$ states form 
4 cycles of length 1, 4 cycles of length 3, 12 cycles of length 5, and 12 cycles of length 15.
We denote this by the cycle polynomial:
$D_3 D_5 = 4 x + 4 x^3 + 12 x^5 + 12 x^{15}$, where $D_3$ and $D_5$ are given in Table 1 and are multiplied according to the rule in Eq. (\ref{Product}).
}
\end{small}
\end{figure}
% - - - - - - - - - - - - - - - - - - - - - - - - - - - - - - - - - - - - - - - - - - - - - - - -
\\ \\ \noindent {\sf\textbf{\textcolor{black}{\large 2 Preliminaries}}}
\\ {\sf\textbf{Dynamics of single and multiple loops}}
% - - - - - - - - - - - - - - - - - - - - - - - - - - - - - - - - - - - - - - - - - - - - - - - -
\\
%\emph{\textcolor{purple}{K=1 critical model.}}
The network in the $K=1$ Kauffman model has a particularly simple structure.
It is composed of loops and trees branching off of loops, as shown in the top of Fig. 1.
The trees that branch off of the loops will eventually fall into periodic behavior determined by the loops, so they do not contribute to the number or length of attractors, which are set by the $m$ nodes in loops.
There are four possible Boolean functions with one input: on, off, copy and invert.
For a Kauffman model to be critical, a change to the value of one node must propagate to, on average, precisely one other node.
So in the critical $K=1$ model, all Boolean functions must be either copy or invert.
In the bottom of Fig.~1 we show the various attractors generated by the 3-loop and the 5-loop in the network.
\\ \indent
%\emph{\textcolor{purple}{Even and odd loops.}}
It turns our that the detailed arrangement of the copy and invert Boolean functions does not matter---only the parity of of the number of inverts influences the number and size of attractors.
We call a loop with an even number of inverts an even loop, 
and a loop with an odd number of inverts an odd loop.
An even loop behaves as if all of the Boolean functions are copy.
An odd loops behaves as if all of the Boolean functions are copy apart from one invert.
\\ \indent
%\emph{\textcolor{purple}{Single loop.}}
{\ifredline\color{red}\else\color{black}\fi 
An even loop of size $l$,       indicated by $\{l\}$,       has attractors of length $k$ if $k$ divides $l$.
An odd loop of size $l$,        indicated by $\{\ov{l}\}$,  has attractors of length $2 k$ if and only if $k$ divides $l$ and $l/k$ is odd~\cite{Flyvbjerg88,Drossel05,Fink23PRE,Fink23PRL}.
For example, for an even 6-loop the cycle lengths are 1, 2, 3 and 6, whereas for an odd 6-loop they are 4 and 12.
Let $A x^\nu$ denote $A$ attractors of length $\nu$.
Then we can represent the number and length of attractors in a loop by the primitive cycle polynomial:
$A_1 x^{\nu_1} + A_2 x^{\nu_2} + \ldots$, 
which we call $D_l$ if the loop is even and $D_{\ov{l}}$ if the loop is odd
(we dropped the braces around $l$ and $\ov{l}$ in $D_l$ and $D_{\ov{l}}$ for convenience).
Table 1 give some examples of primitive cycle polynomials.
}
% - - - - - - - - - - - - - - - - - - - - - - - - - - - - - - - - - - - - - - - - - - - - - - - -
\\ \indent
%\emph{\textcolor{purple}{Multiple loops.}}
A network containing multiple loops has attractor lengths that are the least common multiples of the attractor lengths in the individual loops.
The cycle polynomial for multiple loops can be determined from the cycle polynomials for individuals loops by defining an appropriate product between them.
Given $A$ attractors of length $\nu$ and $B$ attractors of length $\xi$, their product is
	\begin{equation*}
		A x^\nu \cdot B x^\xi =  A B \, {\rm gcd}(\nu, \xi) x^{{\rm lcm}(\nu, \xi)}. 
	\end{equation*}
Then the product between two cycle polynomials (primitive or otherwise) is
	\begin{equation}
		\sum_i A_i x^{\nu_i} 
		\cdot
		\sum_j B_j x^{\xi_j} 
		= \sum_{i,j} A_i B_j {\rm gcd}(\nu_i, \xi_j) x^{{\rm lcm}(\nu_i, \xi_j)},
		\label{Product}
	\end{equation}
first introduced in \cite{Fink23PRE}.
For example, the cycle polynomial for the even 3-loop and even 5-loop in Fig. 1 is
	\begin{eqnarray*}
		D_3(x) D_5(x)   &=& (2 x + 2 x^3) (2 x + 6 x^5)				\\
		                &=& 4 x + 4 x^3 + 12 x^5 + 12 x^{15}.
	\end{eqnarray*}
The cycle polynomial for an even 4-loop and an odd 4-loop is
	\begin{eqnarray*}
		D_4(x) D_{\ov{4}}(x)	&= (2 x + x^2 + 3 x^4) (2 x^8) 					  \\
					    &=  4 x^8 + 4 x^8 + 24 x^8 			  \\		
					    &=  32 x^8.	
	\end{eqnarray*}
The cycle polynomials for other combinations of loops are given in the right side of Table 1. 
%	\begin{eqnarray}
%		&&
%		\Bigg( \sum_i f_i \ci x_i \Bigg)
%		\Bigg( \sum_j g_j \ci y_j \Bigg)
%		\Bigg( \sum_k h_k \ci z_k \Bigg) \ldots  \label{product} \\
%		&& \qquad = \!\! \sum_{i,j,k,\ldots} \!\! (f_i g_j h_k \ldots) \frac{x_i y_j z_k \ldots}{{\rm lcm}(x_i, y_j, z_k, \ldots)} \ci  {\rm lcm}(x_i, y_j, z_k, \ldots). \nonumber
%	\end{eqnarray}
% - - - - - - - - - - - - - - - - - - - - - - - - - - - - - - - - - - - - - - - - - - - - - - - -
\\ \\ \noindent
{\sf\textbf{Attractor number and attractor length}} \\
There are two main quantities of interest in the critical Kauffman model.
One is the number of attractors $c$, and the other is the mean attractor length $\ov{A}$.
For the critical Kauffman model with connectivity one, these two quantities are intimately related.
%$c \, \ov{A} = 2^m$, where $m$ is the number of nodes in loops.
Let $L$ be the set of loops in the network and
{\ifredline\color{red}\else\color{black}\fi
$m$ be the number of nodes in loops.
}
Writing the cycle polynomial of the network as $\sum_i A_i x^{\nu_i}$---this being the product of the primitive cycle polynomials for all of the loops---the number of attractors is
	\begin{eqnarray*}
		c(L) = \sum_{i} \nu_i,
	\end{eqnarray*}
and the mean attractor length is	
	\begin{eqnarray}
		\overline{A}(L) 	&=& \sum_{i} \nu_i A_i \Big/  \sum_{i} \nu_i  	\nonumber \\
				        %&=& \sum_{i} \nu_i A_i \Big/ c(L) 			\nonumber \\
					    &=& 2^m \Big/ c(L),   \label{ConservationLaw}
	\end{eqnarray}
since all $2^m$ states of the loop nodes belong to attractors.
In other words, for a given set of even and odd loops, 
the number of attractors and the mean attractor length satisfy the conservation law $c(L) \overline{A}(L) = 2^m$.
\\ \indent
%\emph{\textcolor{red}{Averaging.}}
Ultimately, we are interested in $c(L)$ and $\overline{A}(L)$ averaged over the ensemble of directed networks with $K=1$ and the distribution over loop parities.
We denote these quantities as
\begin{eqnarray*}
    c(N) = \sum_L c(L)\, P(L\mid N)
    \quad {\rm and} \quad
    \overline{A}(N) = \sum_L \overline{A}(L)\, P(L\mid N).
\end{eqnarray*}
{\ifredline\color{red}\else\color{black}\fi
where $P(L\mid N)$ is the distribution over network configurations and parities for a fixed number of nodes $N$.
}
It is also useful to consider $c(L)$ and $\overline{A}(L)$ averaged over those networks with fixed number of nodes in loops $m$.
%, where $m = l_1 + l_2 + \dots$.
We denotes these as
\begin{eqnarray}\label{eq:maverage}
    c(m) = \sum_L c(L)\, P(L\mid m) 
    \quad {\rm and} \quad
    \overline{A}(m) = \sum_L \overline{A}(L)\, P(L\mid m),
\end{eqnarray}
where we note that $P(L\mid m) = P(L\mid N,m)$.
The two ensemble averages are related by
\begin{equation}\label{eq:mtoN}
    c(N) = \sum_{m=1}^N c(m)\, P(m\mid N)
    \quad {\rm and} \quad
    \overline{A}(N) = \sum_{m=1}^N \overline{A}(m)\, P(m\mid N)
\end{equation}
{\ifredline\color{red}\else\color{black}\fi 
where $P(m\mid N)$ is the distribution over the number of nodes in loops at fixed total number of nodes $N$.}
We will investigate these quantities using insights from number theory in the two sections that follow.
% - - - - - - - - - - - - - - - - - - - - - - - - - - - - - - - - - - - - - - - - - - - - - - - -
\begin{table*}[t!]
\noindent
\begin{small}
\begin{flushright}
\begin{tabular*}{1\columnwidth}{@{\extracolsep{\fill}}llllrll}
\emph{Even loop} 				    & &\emph{Odd loop} 		  	          & & \multicolumn{3}{c}{\emph{Multiple even loops}\hspace{60pt} } 	   \\
$D_1 = 2 x$ 					    & & $D_{\ov{1}} = x^2$ 		          & & $D_1 D_7$           &=& $4 x + 36 x^7$						   \\ 
$D_2 = 2 x + x^2$ 				    & & $D_{\ov{2}} = x^4$			      & & $D_2 D_6$ 	      &=& $4 x + 6 x^2 + 4 x^3 + 38 x^6$		   \\
$D_3 = 2 x + 2 x^3$ 			    & & $D_{\ov{3}} = x^2 + x^6$	      & & $D_3 D_5$ 	      &=& $4 x + 4 x^3 + 12 x^5 + 12 x^{15}$	   \\ 
$D_4 = 2 x + x^2 + 3 x^4$           & & $D_{\ov{4}} = 2 x^8$		      & & $D_4 D_4$ 	      &=& $4 x + 6 x^2 + 60 x^4$			   	   \\
$D_5 = 2 x + 6 x^5$ 			    & & $D_{\ov{5}} = x^2 + 3 x^{10}$	  & & $D_2 D_3 D_3$       &=& $8 x + 4 x^2 + 40 x^3 + 20 x^6$ 	       \\
$D_6 = 2 x + x^2 + 2 x^3 + 9 x^6$   & & $D_{\ov{6}} = x^4 + 5 x^{12}$	  & & $D_2 D_2 D_4$ 	  &=& $8 x + 28 x^2 + 48 x^4$	 	           \\
$D_7 = 2 x + 18 x^7$ 			    & & $D_{\ov{7}} = x^2 + 9 x^{14}$	  & & $D_2^4$             &=& $16 x + 120  x^2$			     	       \\
$D_8 = 2 x + x^2 + 3 x^4 + 30 x^8$  & & $D_{\ov{8}} = 16 x^{16}$	      & & $D_1^8$             &=& $256 x$		
\end{tabular*}
\end{flushright}
\end{small}
\caption{
{\bf Cycle polynomials for single and multiple loops.} 
The primitive cycle polynomials $D_l(x)$ and $D_{\ov{l}}(x)$ indicate the number and length of attractors in an even and odd parity loop of size $l$, respectively.
For example, $D_3(x) = 2 x + 2 x^3$ reads as two cycles of length one and two cycles of length three.
The cycle polynomial for two loops is given by the product of the individual primitive cycle polynomials, where the product is defined by Eq. (\ref{Product}).
Note that we use the shorthand $D_2^4 \equiv D_2 D_2 D_2 D_2$, and so on.
}
\end{table*}
% - - - - - - - - - - - - - - - - - - - - - - - - - - - - - - - - - - - - - - - - - - - - - - - -
\\ \\ \noindent {\sf\textbf{\textcolor{black}{\large 3 Attractor number and Landau's function}}} \\
The first key quantity in the critical Kauffman model is the number of attractors.
Rather than calculate it directly, we take an indirect approach.
We use the conservation law $\ov{A}(L) = 2^m/c(L)$ in Eq. (\ref{ConservationLaw}) to convert an upper bound on the mean attractor length $\ov{A}(L)$ into a lower bound on the the number of attractors $c(L)$.
\\ \indent
We can bound $\ov{A}(L)$ from above by calculating the largest possible attractor length.
Consider all ways of partitioning $m$ nodes into a collection of loops $L = \{l_1, l_2, \ldots\}$, 
where $l_1 + l_2 + \ldots = m$.
For $m = 8$, some of the 22 partitions are shown in the right of Table 1.
From the definition of the product in Eq. (\ref{Product}), we see that
when all of the loops are even, the longest attractor length is the least common multiple of the loop sizes.
When one or more of the loops is odd, the longest attractor length is double this.
\\ \indent
What is the maximum value of the lcm of the partitions of $m$?
This is precisely the definition of Landau's function $g(m)$ (OEIS A000793~\cite{Sloane}), 
which we briefly review in the subsection below.
Thus the mean attractor length satisfies
	\begin{equation*}
		\overline{A}(L) < 2 g(m).
	\end{equation*}
Inserting the bound on $g(m)$ from Eq. (\ref{MS}) below, we find
	\begin{equation}
		\overline{A}(L) < 2^{1.52 \sqrt{m \ln m}+1}.
		\label{AUpperBound}
	\end{equation}
Inserting this into Eq. (\ref{ConservationLaw}) and averaging at fixed $m$ with Eq.~(\ref{eq:maverage}) gives a lower bound on the number of attractors,
	\begin{equation}
		c(m) > 2^{m - 1.52 \sqrt{m \ln m}-1}.
    \label{cmbound}
	\end{equation}
% - - - - - - - - - - - - - - - - - - - - - - - - - - - - - - - - - - - - - - - - - - - - - - - -
\begin{table}[t!]
\noindent
\begin{small}
\begin{flushright}
\begin{tabular*}{0.60\columnwidth}{@{\extracolsep{\fill}}rrrl}
        &       & \emph{Landau's}   & \emph{Partition of $m$ with largest} \\ 
$m$     & $s$   & \emph{function} $g(m)$   & \emph{least common multiple} \\ 
1       &       & 1                 & $\{ 1 \}$                         \\    
2       & 1              & 2                 & $\{ 2 \}$                         \\    
3       &               & 3                 & $\{ 3 \}$                         \\    
4       &               & 4                 & $\{ 4 \}$                         \\    
5       & 2              & 6                 & $\{ 2,3 \}$                       \\       
6       &               & 6                 & $\{ 1,2,3 \}$                     \\    
7       &               & 12                & $\{ 3,4 \}$                     \\    
8       &               & 15                & $\{ 3,5 \}$                       \\    
9       &               & 20                & $\{ 4,5 \}$                       \\    
10      & 3              & 30                & $\{ 2,3,5 \}$                     \\       
17      & 4              & 210               & $\{ 2,3,5,7 \}$                   \\       
28      & 5              & 2,310             & $\{ 2,3,5,7,11 \}$                \\       
41      & 6              & 30,030            & $\{ 2,3,5,7,11,13 \}$             \\       
58      & 7             & 510,510           & $\{ 2,3,5,7,11,13,17 \}$          \\   
77      & 8             & 9,699,690         & $\{ 2,3,5,7,11,13,17,19 \}$       \\       
100     & 9             & 232,792,560       & $\{3,5,7,9,11,13,16,17,19 \}$    \\  
129     & 10            & 7,216,569,360     & $\{3,4,5,6,7,11,17,19,26,31 \}$   
%129     & 10            & 7,216,569,360     & $\{ 5, 7, 9, 11, 13, 16, 17, 19, 31 \}$   
%100     & 9             & 232,792,560       & $\{ 3,5,7,9,11,13,16,17,19 \}$    \\  
\end{tabular*}
\end{flushright}
\end{small}
\caption{
{\bf Values of Landau's function and the partitions that generate them.} 
Landau's function $g(m)$ is the maximum value of the least common multiple of the partitions of $m$.
We show this for $m=1$ to 10 as well as when $m$ is the sum of the first $s$ primes, up to $s=10$.
For $s \leq 8$, $g(m)$ is just the product of the first $s$ primes.
But this is not true in general, as $s = 9$ and $s = 10$ show.
%for $s = 9$, ${\rm lcm}(2,3,\ldots,19,23) < {\rm lcm}(9,3,\ldots,19,16)$, 
%But notice that ${\rm lcm}(2,3,5,7,11,13,17,19,23) < {\rm lcm}(9,3,5,7,11,13,17,19,16)$, 
%where we replaced 2 and 23 with 9 and 16.
}
\end{table}
% - - - - - - - - - - - - - - - - - - - - - - - - - - - - - - - - - - - - - - - - - - - - - - - -
\mbox{} \indent 
To re-express this in terms of the total number of nodes $N$, we must average over the distribution of the number of nodes in loops $m$ using Eq.~(\ref{eq:mtoN}).
% In the large $N$ limit, the mean number of loops of size $l$ is $\exp\left(-l^2/(2N)\right)/l$.
In the large $N$ limit, the probability of $m$ of $N$ nodes occurring in loops is $\frac{m}{N}\exp\left(-m^2/(2N)\right)$~\cite{Flyvbjerg88}.
Summing over this, the mean number of nodes in loops $\overline{m}$ is asymptotically $\sqrt{{\pi \over 2} N}$ \cite{Flyvbjerg88,Drossel05}.
Since Eq. (\ref{cmbound}) is convex, by Jensen's inequality we can replace $m$ with its mean, giving
	\begin{equation}\label{ClowerN}
		c(N) > 2^{1.25 \sqrt{N} - \sqrt[4]{N}\sqrt{1.45 \ln N + 0.65}-1}. 
        %= O(2^{1.25 \sqrt{N}}).
	\end{equation}
Taking the logarithm, the leading behavior of the number of attractors is
    \begin{eqnarray}
        \log_2 c(N) > 1.25 \sqrt{N},
    \end{eqnarray}
where we have neglected constant terms and terms proportional to $\sqrt[4]{N} \sqrt{\ln N}$.
% - - - - - - - - - - - - - - - - - - - - - - - - - - - - - - - - - - - - - - - - - - - - - - - -
\\ \\ \noindent
{\sf\textbf{Landau's function and its bound}} \\
Landau's function $g(m)$ is the maximum value of the least common multiple of the partitions of $m$.
It is shown in Table 2 and Fig. 2.
%Partitions into the first $s$ primes are high-water marks for Landau's function for $s \in [1,8]$.
When $m$ is the sum of the first $s \leq 8$ primes, $g(m)$ is just the product of the primes:
	\begin{eqnarray*}
        g(2) &=& 2 \\
        g(2+3) &=& 2 \cdot 3 \\
        g(2+3+5) &=&  2 \cdot 3 \cdot 5,
    \end{eqnarray*}
and so on.
%$g(2) = 2$, $g(5) = 6$, and so on.
But this is not true in general: for $s = 9$, 
    \begin{eqnarray*}
        g(2+3+\ldots+23) &=& 3 \cdot 5 \cdot 7 \cdot 9 \cdot 11 \cdot 13 \cdot 16 \cdot 17 \cdot 19 \\
        &>& 2 \cdot 3 \cdot 5 \cdot 7 \cdot 11 \cdot 13 \cdot 17 \cdot 19 \cdot 23,
    \end{eqnarray*}
%$m=100$ and ${\rm lcm}(2,3,5,7,11,13,17,19,23) < {\rm lcm}(9,3,5,7,11,13,17,19,16)$, 
%But notice that ${\rm lcm}(2,3,5,7,11,13,17,19,23) < {\rm lcm}(9,3,5,7,11,13,17,19,16)$, 
where 2 and 23 are replaced with 9 and 16.
% - - - - - - - - - - - - - - - - - - - - - - - - - - - - - - - - - - - - - - - - - - - - - - - -
\begin{figure}[t!]
\noindent
      \epsfxsize = \columnwidth \epsfbox{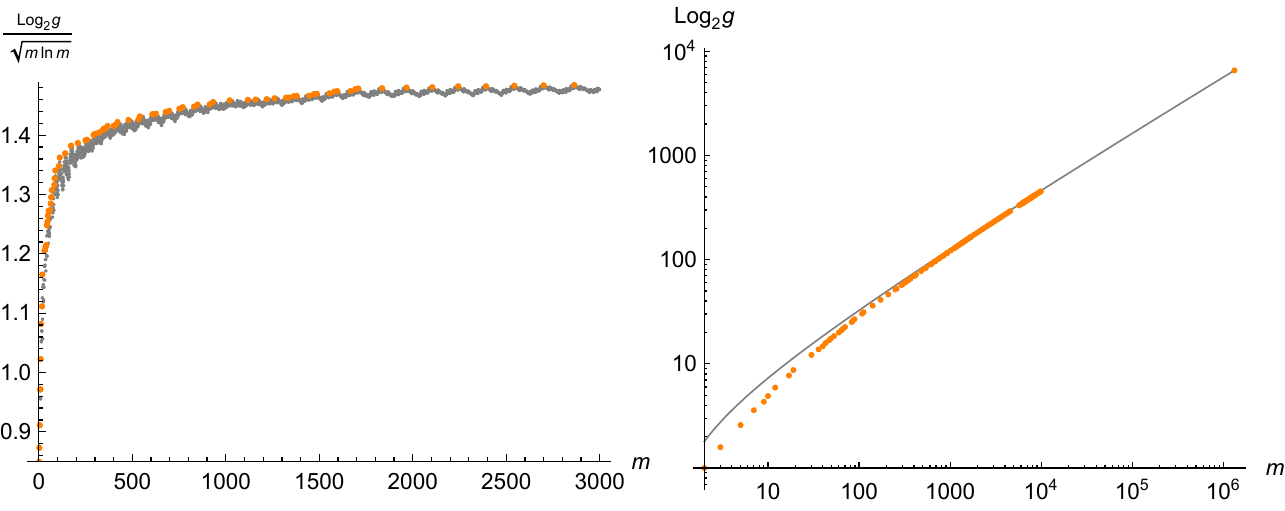}
\begin{small}
\caption{
{\bf Bound on Landau's function.} 
{\bf Left.} 
The smaller points show the first 3000 values of $\log_2 g(m)/\sqrt{m \ln m}$,
and the larger points (orange online) show their running maxima.
Massias showed that there are only 378 such maxima, the largest occurring at $m^* =$ 1,319,766.
{\bf Right.} 
The line is $1.52\sqrt{m \ln m}$, and the points are the first 122 running maxima (the ones with position up to $10^4$), as well as the last. 
Thus $\log_2 g(m) \leq 1.52\sqrt{m \ln m}$, with equality at $m^*$.
}
\end{small}
\end{figure}
% - - - - - - - - - - - - - - - - - - - - - - - - - - - - - - - - - - - - - - - - - - - - - - - -
\\ \indent
In the early 1900s, Landau proved that $\ln g(m)$ asymptotically approaches $\sqrt{m \ln m}$.
More recently, Massias \cite{Massias85} proved
	\begin{equation}
        g(m) \leq 2^{1.52 \sqrt{m \ln m}}.
        \label{MS}
    \end{equation}
In particular, Massias calculated the running maxima of $g(m)/\sqrt{m \ln m}$, 
and showed that there are precisely 378 such maxima (OEIS A103635~\cite{Sloane}).
The first 79 of these are shown in Fig. 2 left.
The position of the last maxima is $m^*=$ 1,319,766, which is where $\ln (g(m))/\sqrt{m \ln(m)}$ reaches its global maximum, namely
	\begin{equation*}
        \frac{\log_2 \left( g(m^*)\right)}{\sqrt{m^* \, \ln m^*}} = 1.519\dots, 
	\end{equation*}
where the prime factorization of $g(m^*)$ is
 	\begin{equation*}
        g(m^*) = 2^9 \cdot 3^6 \cdot 5^4 \cdot 7^3 \cdot 11^3 \cdot (13^2 \cdot \ldots \cdot 43^2) \cdot (47 \cdot \ldots \cdot 4409) \simeq 10^{1972}.
	\end{equation*}
% - - - - - - - - - - - - - - - - - - - - - - - - - - - - - - - - - - - - - - - - - - - - - - - -
\mbox{} \\ \noindent {\sf\textbf{\textcolor{black}{\large 4 Attractor length and Mertons' theorems}}} \\
% - - - - - - - - - - - - - - - - - - - - - - - - - - - - - - - - - - - - - - - - - - - - - - - -
The second key quantity in the critical Kauffman model is the mean attractor length. 
For a given network, the mean attractor length $\ov{A}$ has a lower bound that is proportional to the maximum attractor length \cite{Drossel05},
with a constant of proportionality that we call $\zeta$.
When all of the loops are even, the maximum attractor length is the least common multiple of the loop sizes.
When one or more of the loops is odd, it is twice this. 
To write down $\ov{A}$, let $\sigma$ be a binary vector of length $N$, where $\sigma_l=1$ if at least one loop of length $l$ occurs and $\sigma_l=0$ otherwise.
Given the distribution over the $\sigma$s,
    \begin{eqnarray*}
        \ov{A}    & > \zeta \sum_{\sigma}P(\sigma) \, {\rm lcm}\left(1^{\sigma_1}, 2^{\sigma_2}, \dots, N^{\sigma_N}\right).
    \end{eqnarray*}
Dropping all but the prime loops gives us a weaker bound, but one in which we no longer have to worry about the lcm:
    \begin{eqnarray}
        \ov{A} & >  \zeta \sum_{\sigma}P(\sigma) \,\prod_{p \leq N}\,  p^{\sigma_p},
        \label{NA}
    \end{eqnarray}
where the product is over all primes $p$ less than $N$.
\\ \indent
The key obstacle to calculating this quantity is that the presence of loops is not independent.
However, Drossel and Greil~\cite{Drossel05} argue that below a critical loop length $l^*= \sqrt{\epsilon N}$, where $\epsilon$ is a small constant, loops of different sizes are independent and are Poisson distributed with mean $1/l$. 
Therefore the probability of finding at least one loop of size $l < l^*$ is $1-e^{-1/l}$,
which to second order is $1/l - 1/(2 l^2)$.
Using Step 1 below, which gives the expected value of the product of randomly selected primes,
we can write Eq. (\ref{NA}) as
    \begin{eqnarray}
        \ov{A} > \zeta \, 2^{l^*/\ln l^*} \prod_{p < l^*} \textstyle \left(1 - \frac{3}{4 p} \right).
        \label{QA}
    \end{eqnarray}
Using Step 2 below, we can compute the Euler product in Eq. (\ref{QA}), which decreases slowly with $l^*$:
\begin{eqnarray*}
    \ov{A} > \zeta \, 2^{l^*/\ln l^*} \frac{0.710}{(\ln l^*)^{3/4}}.
\end{eqnarray*}
Substituting in $l^*= \sqrt{\epsilon N}$ and taking the logarithm, the leading behavior of the mean attractor length is
\begin{eqnarray*}
    \log_2 \ov{A} > \frac{\sqrt{\epsilon N}}{\ln \sqrt{\epsilon N}},
\end{eqnarray*}
where we have neglected constant terms and terms proportional to $\ln \ln \sqrt{\epsilon N}$.
% - - - - - - - - - - - - - - - - - - - - - - - - - - - - - - - - - - - - - - - - - - - - - - - -
\\ \\ \noindent
{\sf\textbf{Step 1: Primes in a jar}} \\
Consider the following problem, which has a surprisingly simple solution.
Put the number 1 and the first $s$ primes in a jar, each with probability equal to the inverse of the number itself.
What is the expected value of the product of the numbers?
For example, for $s=3$, if we put the numbers 1, 2, 3 and 5 in a jar with probability 1, 1/2, 1/3 and 1/5, 
the expected value of their product is 9/2.
\\ \indent
More generally, label the primes 2, 3, 5, \ldots as $p_1, p_2, p_3, \ldots$,
and let $u_i$ be the probability that prime $p_i$ is in the jar, with $v_i = 1 - u_i$.
Then the expected value $z$ of the product can be had by summing over the $2^s$ configurations, namely,
\begin{eqnarray*}
    z   &=& \!\!\!\!\! \sum_{\sigma_1, \ldots, \sigma_s = 0}^1 \!\!\!\!\! p_1^{\sigma_1} u_1^{\sigma_1} v_1^{1-\sigma_1} \ldots p_s^{\sigma_s} u_s^{\sigma_s} v_s^{1-\sigma_s} \\
    &=& \prod_{i=1}^s (p_i u_i + v_i).                                      
\end{eqnarray*}
With $u_i = 1/p_i$, which is the probability used in our example above,
    \begin{eqnarray*}
        z   &=& \prod_{i=1}^s {\textstyle \left(2 - \frac{1}{p_i}\right)}     \\
            &=& 2^s \prod_{i=1}^s \textstyle \left(1 - \frac{1}{2 p_i} \right).                           
    \end{eqnarray*}
To obtain a lower bound, we expand $p_i = 1-e^{-\frac{1}{p_i}}$ and keep terms to first order in $1/p_i$.
Then
\begin{eqnarray}
 z       &\ge& 2^s \prod_{i=1}^s \textstyle \left(1 - \frac{3}{4 p_i} \right).  
    \label{OA}
\end{eqnarray}
Instead of considering the first $s$ primes, we can just as well consider primes less than some cutoff $l^*$.
The number of primes less than or equal to $l^*$ is asymptotically $l^* / \ln l^*$,
which is a lower bound when $l^*> 17$ \cite{Tenenbaum}.
Then Eq. (\ref{OA}) becomes
\begin{eqnarray*}
    z   &\ge& 2^{l^*/\ln l^*} \prod_{p < l^*} \textstyle \left(1 - \frac{3}{4 p} \right),
\end{eqnarray*}
where the product is over all primes $p$ less than $l^*$.
% - - - - - - - - - - - - - - - - - - - - - - - - - - - - - - - - - - - - - - - - - - - - - - - -
\\ \\ \noindent {\sf\textbf{Step 2: Generalizing Merten's third theorem}} \\
Mertens' theorems are three results in number theory that generalise familiar sums and products of integers to sums and products of primes~\cite{Tenenbaum}.
For example, the second theorem states
\begin{eqnarray*}
    \lim_{l^*\to\infty} \Bigg(\sum_{p\le l^*} {\textstyle \frac{1}{p}} - \ln \ln l^*\Bigg) = M =0.261\dots,
\end{eqnarray*}
where $M$ is the Meissel-Mertens constant and the sum is over primes.
This is the analogue for primes of the more familiar definition of the Euler-Mascheroni constant $\gamma$:
\begin{eqnarray*}
        \lim_{l^*\to\infty} \Bigg(\sum_{n\le l^*} {\textstyle \frac{1}{n}} - \ln l^*\Bigg) = \gamma=0.577\dots,
\end{eqnarray*}
where the sum is over integers.
The reduction in the rate of growth from $\ln l^*$ to $\ln \ln l^*$ is the result of the smaller number of terms in the sum over primes.
\\ \indent
The Euler product at the end of Step 1 is reminiscent of Merten's third theorem~\cite{Tenenbaum}, which states
\begin{equation*}
    \lim_{l^*\to\infty} \ln l^* \prod_{p\le l^*} \left(\textstyle{1 - \frac{1}{p}}\right) = e^{-\gamma}.
\end{equation*}
Here, we generalize this to
\begin{equation*}\label{eqn:genmert}
    \lim_{l^*\to\infty} \,(\ln l^*)^x \prod_{p\le l^*} \left(\textstyle{1 - \frac{x}{p}}\right) = e^{-x(\gamma + D(x)) },
\end{equation*}
where $D(x)$ is a constant defined below.
To do so, we start by writing
\begin{eqnarray}
    \lim_{l^*\to\infty} (\ln l^*)^x
    \prod_{p\le l^*} \left(\textstyle{1 - \frac{x}{p}}\right) 
    = 
    \lim_{l^*\to\infty} (\ln l^*)^x
    \exp\Bigg(\sum_{p\le l^*}  \ln\left(\textstyle{1 - \frac{x}{p}}\right)\Bigg) \nonumber \\
    \qquad =
    \lim_{l^*\to\infty}
    \exp\Bigg(-x\Big(\sum_{p\le l^*}{\textstyle \frac{1}{p}} - \ln\ln l^*\Big) +\sum_{p\le l^*}  \left(\ln\left(\textstyle{1 - \frac{x}{p}}\right) + \textstyle \frac{x}{p}\right)\Bigg).
    \label{RA}
\end{eqnarray}
The Meissel-Mertens constant can be expressed in two equivalent ways:
    \begin{eqnarray*}
        M = \lim_{l^*\to\infty} \Bigg(\sum_{p\le l^*} {\textstyle \frac{1}{p}} - \ln \ln l^*\Bigg)
        = \gamma + \sum_{p} \textstyle{\left(\ln\left(1-\frac{1}{p}\right) + \frac{1}{p}\right)}.
    \end{eqnarray*}
Inserting this into Eq. (\ref{RA}) gives
\begin{eqnarray*}
    \lim_{l^*\to\infty} (\ln l^*)^x \prod_{p\le l^*} \left(\textstyle{1 - \frac{x}{p}}\right) \\
    \qquad = \exp\Bigg( -x \gamma
    -x\Bigg(\sum_{p} \left(\textstyle{\ln\left(1-\frac{1}{p}\right) + \frac{1}{p}}\right) \Bigg)
    + \sum_{p}  \left(\ln\left(\textstyle{1 - \frac{x}{p}}\right) + \textstyle \frac{x}{p}\right)\Bigg)\\
    \qquad = \exp\Bigg(-x \gamma -x \Bigg(\sum_{p} \ln\frac{\textstyle{1-\frac{1}{p}}}{\textstyle{\big(1-\frac{x}{p}\big)^{\frac{1}{x}}}}
    \Bigg)\Bigg).
\end{eqnarray*}
Thus we arrive at our desired generalization,
\begin{equation*}
    \lim_{l^*\to\infty} \,(\ln l^*)^x\prod_{p \le l^*} \left(\textstyle{1 - \frac{x}{p}}\right) = e^{-x(\gamma + D(x))},
\end{equation*}
where 
\begin{equation*}
D(x) = \sum_{p} \ln\frac{\textstyle{1-\frac{1}{p}}}{\textstyle{\Big(1-\frac{x}{p}\Big)^{\frac{1}{x}}}}
\end{equation*}
is a constant and, in particular, $D(3/4) = -0.102\dots$.
% - - - - - - - - - - - - - - - - - - - - - - - - - - - - - - - - - - - - - - - - - - - - - - - -
\begin{figure}[b!]
    \centering
    \includegraphics[width=\columnwidth]{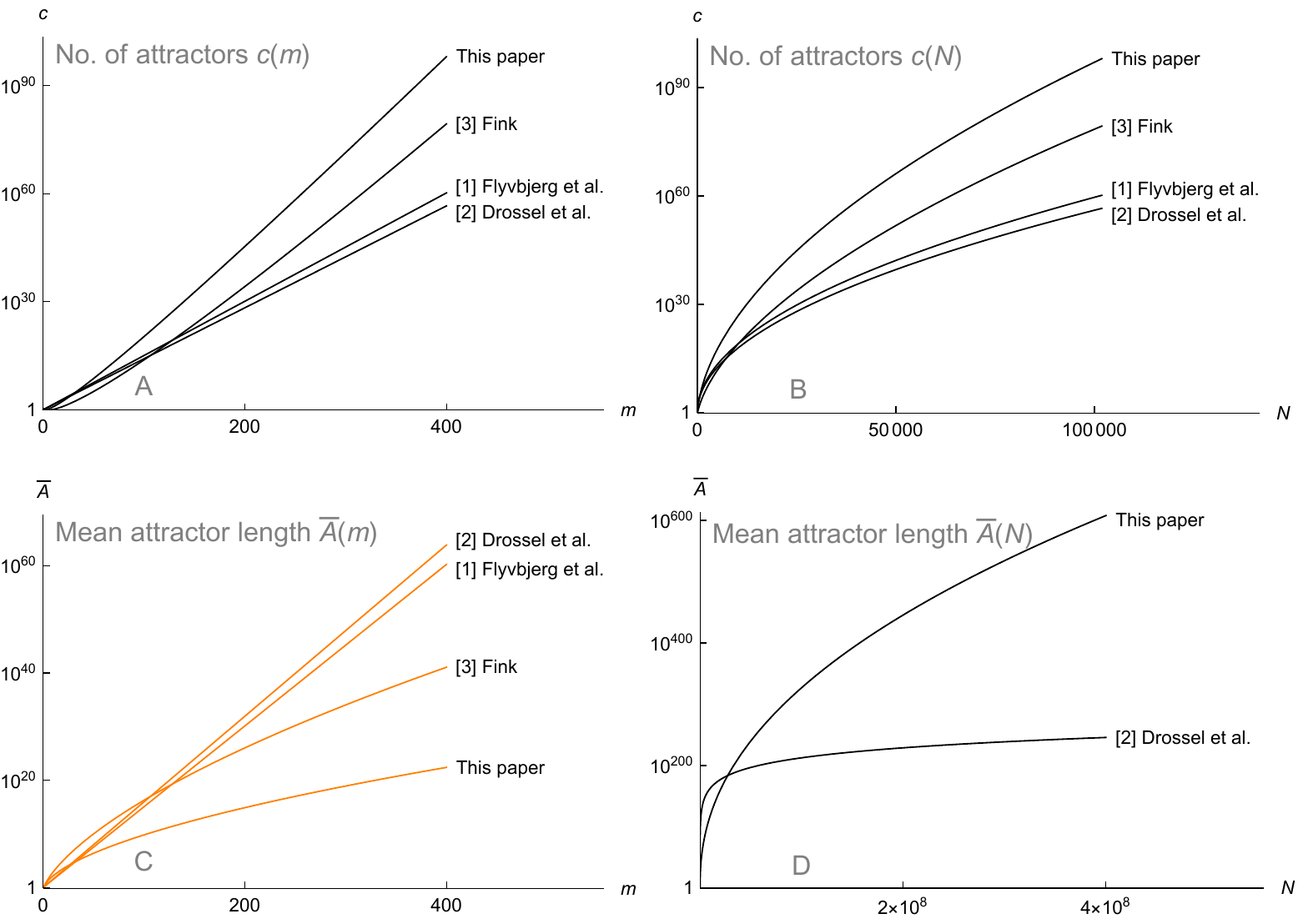}
    \vspace{-0.15in}
    \caption{
    \textbf{Bounds on $\boldsymbol{c}$ and $\boldsymbol{\overline{A}}$.} 
    We compare our results with the known analytic results for the number of attractors $c$ and the mean average attractor length $\overline{A}$,
    as a function of both the number of nodes in loops $m$ and the size of the network $N$.
    Lower bounds are black and upper bounds are grey (orange online).
    \textbf{A}. 
    Our lower bound on the number of attractors $c(m)$ exceeds those given in 
    \cite{Flyvbjerg88,Drossel05,Fink23PRE}.
    \textbf{B}. 
    The same is true for the number of attractors $c(N)$.
    \textbf{C}. 
    We did not calculate a lower bound on the mean attractor length $\overline{A}(m)$,
    but we can translate the lower bounds on $c(m)$ into upper bounds on $\overline{A}(m)$ using $c(m) \overline{A}(m) = 2^m$.
    \textbf{D}. 
    Our lower bound on the mean attractor length $\overline{A}(m)$ exceeds that given in 
    \cite{Flyvbjerg88}, where we have taken any constants to be unity.
    For large enough $N$, the constants are immaterial.
    }
    \label{ColorFitnessComparison}
\end{figure}
% - - - - - - - - - - - - - - - - - - - - - - - - - - - - - - - - - - - - - - - - - - - - - - - -
\\ \\ \noindent {\sf\textbf{\textcolor{black}{\large 5 Discussion}}} \\
% - - - - - - - - - - - - - - - - - - - - - - - - - - - - - - - - - - - - - - - - - - - - - - - -
%\emph{\textcolor{red}{Four quantities.}}
Insights from number theory have enabled us to improve on the best-known bounds on the number of attractors $c$ and the mean attractor length $\overline{A}$.
In Fig.~3 we compare our results with the known analytic results.
We express $c$ and $\overline{A}$ in terms of both the number of nodes in loops $m$ and the size of the network $N$.
Each of these four functions
is shown in a different panel in Fig.~3, and calculating each presents its own challenges.
We show lower bounds as black and upper bounds as grey (orange online).
When presenting results below, we retain terms to lowest order in the exponent, apart from when this is $m$, and the deviation from $m$ is of interest. 
% - - - - - - - - - - - - - - - - - - - - - - - - - - - - - - - - - - - - - - - - - - - - - - - -
\\ \indent
%\emph{\textcolor{red}{$c(m)$.}}
The number of attractors $c(m)$ is shown in Fig.~3A.
We find that $c(m) > 2^{m - 1.52 \sqrt{m \ln m}}$.
This is a faster growth rate than
$2^{0.5 m}$ in \cite{Flyvbjerg88},
$2^{0.47 m}$ in \cite{Drossel05}, and
$2^{m - \sqrt{2m}\log_2 \sqrt{2m}}$ in \cite{Fink23PRE}.
Notice that for our result here and the result in \cite{Fink23PRE}, the exponent is asymptotic to $m$.
% - - - - - - - - - - - - - - - - - - - - - - - - - - - - - - - - - - - - - - - - - - - - - - - -
\\ \indent
%\emph{\textcolor{red}{$c(N)$.}}
The number of attractors $c(N)$ is shown in Fig.~3B.
We can convert our bound on $c(m)$ to bounds on $c(N)$ by using Jensen's inequality. 
Since $2^{m - 1.52 \sqrt{m \ln m}}$ is convex, the bound holds on replacing $m$ with its mean $\overline{m}$.
We obtain $c(N) > 2^{1.25 \sqrt{N}}$,
which is faster than 
$2^{0.63 \sqrt{N}}$ in \cite{Flyvbjerg88} and
$2^{0.59 \sqrt{N}}$ in \cite{Drossel05}, and similar to
$2^{1.25 \sqrt{N}}$ in \cite{Fink23PRE}, which was the result of a much lengthier calculation.
% - - - - - - - - - - - - - - - - - - - - - - - - - - - - - - - - - - - - - - - - - - - - - - - -
\\ \indent
%\emph{\textcolor{red}{$\overline{A}(m)$.}}
The mean attractor length $\overline{A}(m)$ is shown in Fig.~3C.
Historically $\overline{A}$ has proved to be more difficult to quantify than the attractor number.
We do not have a lower bound on $\overline{A}(m)$, but we converted the lower bounds on $c(m)$ to upper bounds on $\overline{A}(m)$ using the conservation law in Eq.~(\ref{ConservationLaw}).
Our upper bound, $\overline{A}(m) < 2^{1.52 \sqrt{m \ln m}}$, is slower than
$2^{0.5 m}$ in \cite{Flyvbjerg88},
$2^{0.53 m}$ in \cite{Drossel05}, and
$2^{\sqrt{2m}\log_2 \sqrt{2m}}$ in \cite{Fink23PRE}.
% - - - - - - - - - - - - - - - - - - - - - - - - - - - - - - - - - - - - - - - - - - - - - - - -
\\ \indent
%\emph{\textcolor{red}{$\overline{A}(N)$.}}
The mean attractor length $\overline{A}(N)$ is shown in Fig.~3D.
The difficulty in calculating a lower bound on $\overline{A}(N)$ comes from correlations in the presence of loops of different sizes.
To sidestep this issue, we rely on a claim by Drossel \emph{et al.}~\cite{Drossel05} that these correlations are negligible below a critical loop size $\sqrt{\epsilon N}$, where $\epsilon$ is a small constant.
% To calculate a lower bound on $\overline{A}(N)$, we had to contend with the distribution of loops in the network, which are non-negligibly correlated above loop size $\sqrt{\epsilon N}$~\cite{Drossel05}, where $\epsilon$ is a small constant.
By neglecting contributions above this size and invoking Mertens' theorems, we find that $\overline{A}(N) > 2^{\sqrt{\epsilon N}/ \ln \sqrt{\epsilon N}}$.
The only other analytic result for $\overline{A}(N)$ is from \cite{Drossel05}:
to leading order in the exponent, $\overline{A}(N) > 2^{\alpha (\ln N)^2}$, where $\alpha$ is a constant.
For large enough $N$, the constants $\epsilon$ and $\alpha$ are immaterial, and our bound grows faster than the one in \cite{Drossel05}.
For the plots in Fig.~3D we set the constants to 1.
% - - - - - - - - - - - - - - - - - - - - - - - - - - - - - - - - - - - - - - - - - - - - - - - -
\\ \indent 
%\emph{\textcolor{red}{Discussing consequences.}}
We can also gain some sense of the quality of our lower bound on the mean attractor length $\ov{A}(N)$ by comparing it with our upper bound on $\ov{A}(m)$.
To leading order in the exponent,  
the lower bound on $\log_2 \ov{A}(N)$ is proportional to $\sqrt{N}\big/\log \sqrt{N}$, whereas the upper bound on $\log_2 \ov{A}(m)$ is proportional to $\sqrt{m}$. 
Were $m$ sufficiently concentrated about its mean, $\overline{m} \sim \sqrt{{\pi \over 2} N}$, this would imply that the upper {\ifredline\color{red}\else\color{black}\fi bound} of $\ov{A}(N)$ is proportional to $\sqrt[4]{N}$.
But this cannot be right, because the two bounds would cross.
Consequently, Jensen's inequality, which we have used to translate the bound on $\ov{A}$ from $m$ to $N$, is unlikely to give a tight bound. 
% - - - - - - - - - - - - - - - - - - - - - - - - - - - - - - - - - - - - - - - - - - - - - - - -
\\ \indent
{
\ifredline\color{red}\else\color{black}\fi
So much for our comparison with the literature.
We conclude with some remarks on the connection to number theory and the scope for further work.
\\ \indent
To a reader from physics, it may seem that these links to number theory are incidental, but we believe they hint at a deeper connection.
The growth of the least common multiple of the loop lengths with system size is the central quantity in understanding the Kauffman model.
This is closely related to Chebyshev's second function---the least common multiple of the integers from 1 to $n$---which is a central object of study in analytic number theory.
Exploiting this connection may put within reach the exact averaging over loop length distributions.
In fact, number theory provides a natural language for the $K=1$ critical Kauffman model, as many properties of the model can be naturally described by Dirichlet convolutions.
It appears number theory is at the heart of this problem, and we see many opportunities for further research in this direction.
%But number theory the heart of the $K=1$ Kauffman model.
}
\\ \indent
{
\ifredline\color{red}\else\color{black}\fi
The connection to number theory also critically depends on the update schedule.
Updating all nodes simultaneously preserves intricate periodic patterns on loops, as well as their relative phases.
This leads to the product rule in Eq.~(\ref{Product}) and the prominence of the lcm of loop lengths.
While most research on the Kauffman model assumes synchronous updates, shifting to asynchronous updates significantly reduces the number of attractors~\cite{Greil05,Klemm05}.
Our results are thus limited to the synchronous case.
}
\\ \indent
{\ifredline\color{red}\else\color{black}\fi
There are many open questions concerning the Kauffman model.
The $K=1$ version plays a special role in understanding it because it is exactly solvable~\cite{Flyvbjerg88}.
New approaches to solving it, like the one presented here, reveal additional insights, suggesting techniques for more complex models which cannot be solved exactly.
One obvious target is the original $K=2$ model.
Here, the analogs of loops for $K=1$ are called relevant components.
The cycle polynomials for relevant components multiply in the same way described in Eq. (\ref{Product}) to give the cycle polynomial of the whole network.
However, understanding the exact behavior for $K=2$ relevant components is much harder than for $K=1$.
A stepping stone in the $K=2$ direction would be to consider simple network structures only slightly more complex than loops, such as two loops with a cross-link and one loop with an additional internal link \cite{Kaufman05}.
Initial investigations lead us to believe that the cycle polynomials for such structures could also be concisely expressed in number theoretic terms. 
\\ \indent 
While this paper was under review, we were able to use our result involving Landau's function as part of a more ambitious proof that the number of attractors grows as $(2/\sqrt{e})^N$. 
The resulting paper, which is a follow-up to this one, recently appeared in \emph{Physical Review Letters} \cite{Fink23PRL}.
}
% This growth rate for the number of attractors is a significant departure from $2^{1.25 \sqrt{N}}$ derived here.
% It suggests that the growth rate for the mean attractor length may also have a surprise in store, being much less than the best known upper bound.
% - - - - - - - - - - - - - - - - - - - - - - - - - - - - - - - - - - - - - - - - - - - - - - - -
\ifredline
\else
\\ \noindent
{\sf \textbf{Acknowledgements}}\\
T. Fink and F. Sheldon acknowledge support from bit.bio.
\fi


\vspace{20pt}
\begin{thebibliography}{99}
\bibitem{Kauffman69a}   S. Kauffman,                       Homeostasis and differentiation in random genetic control networks,	                Nature 	          {\bf 224}, 5215        (1969).
\bibitem{Kauffman69b}   S. Kauffman,                       Metabolic stability and epigenesis in randomly constructed genetic nets,	            J Theor Bio       {\bf 22},  437 	    (1969).
\bibitem{Aldana03}      M. Aldana, S. Coppersmith, L. P. Kadanoff,  \textit{Boolean dynamics with random couplings} in \textit{Perspectives and Problems in Nolinear Science} (Springer-Verlag, New York, 2003).
\bibitem{Drossel08}     B. Drossel,                        Random Boolean Networks, in Reviews of Nonlinear Dynamics and Complexity,            (Wiley-VCH,               Weinheim,               2008).
\bibitem{Munoz18} 		M. Mu\~noz, 					   Criticality and dynamical scaling in living systems,							        Rev Mod Phys 	  {\bf 90}, 031001  (2018).
\bibitem{Daniels18}     B. Daniels et al.,                 Criticality distinguishes the ensemble of biological regulatory networks,            Phys Rev Lett     {\bf 121}, 138102 (2018).
\bibitem{Derrida86a}    B. Derrida, D. Stauffer,           Phase transitions in two-dimensional Kauffman cellular automata,                     Europhys Lett     {\bf 2}, 10 (1986).
\bibitem{Derrida86b}    B. Derrida, H. Flyvbjerg,          Multivalley structure in Kauffman's model: analogy with spin glasses,                J Phys A          {\bf 19}, L1003 (1986).
\bibitem{Flyvbjerg88}	H. Flyvbjerg, N. Kj\ae r,          Exact solution of Kauffman's model with connectivity one, 						    J Phys A 		  {\bf 21}, 1695 	  (1988).
\bibitem{Drossel05}		B. Drossel, T. Mihaljev, F. Greil, Number and length of cycles in a critical Kauffman model with connectivity one, 	Phys Rev Lett 	  {\bf 94}, 088701  (2005).
\bibitem{Fink23PRE}		T. Fink,		                   Exact dynamics of the critical Kauffman model with connectivity one, 				arXiv:2302.05314.
\bibitem{Fink23PRL}		T. Fink, F. Sheldon.		       Number of cycles in the critical Kauffman model is exponential, 		            Phys Rev Lett, 	  {\bf 131}, 267402 (2023).
\bibitem{Socolar03}		J. Socolar, S. Kauffman,           Scaling in ordered and critical random Boolean networks,						        Phys Rev Lett 	  {\bf 90}, 068702  (2003).
\bibitem{Bastolla98a}   U. Bastolla, G. Parisi,	           The modular structure of Kauffman networks,                                          Physica D         {\bf 15},  219 	    (1998).
\bibitem{Bastolla98b}   U. Bastolla, G. Parisi,            Relevant elements, magnetization and dynamic properties in Kauffman networks: a numerical study,       Physica D,  {\bf 115}, 203         (1998).
\bibitem{Troein03}		B. Samuelsson, C. Troein,          Superpolynomial growth in the number of cycles in Kauffman networks,		        Phys Rev Lett 	  {\bf 90}, 098701       (2003).
\bibitem{Peixoto10} 	T. Peixoto,						   Redundancy and error resilience in Boolean networks,						            Phys Rev Lett 	  {\bf 104}, 048701      (2010).
\bibitem{Sloane} 		N. Sloane, editor, 				   On-Line Encyclopedia of Integer Sequences, 								            https://oeis.org, 				  2023.
\bibitem{Massias85}     J.-P. Massias,                     Majoration explicite de l’ordre maximum d’un \'{e}l\'{e}ment du groupe sym\'{e}trique,  Ann Fac Sci Toulouse Math {\bf 6},  269     (1984).
\bibitem{Tenenbaum}     G. Tenenbaum, \textit{Introduction to Analytic and Probabilistic Number Theory}                                         (Cambridge University Press, Cambridge, 1995).
\bibitem{Greil05}       F. Greil, B. Drossel, 			   Dynamics of critical Kauffman networks under asynchronous stochastic update,			Phys Rev Lett 	  {\bf 95},  048701 	 (2005).
\bibitem{Klemm05}       K. Klemm, S. Bornholdt, 		   Stable and unstable cycles in Boolean networks,						            Phys Rev E 	      {\bf 72},  055101(R) 	 (2005).
\bibitem{Kaufman05}     V. Kaufman, B. Drossel, 		   On the properties of cycles of simple Boolean networks,						        Eur Phys J B 	  {\bf 43},  115		 (2005).
\end{thebibliography}
\end{document}